\documentclass[preprintnumbers,amsmath,amssymb,floatfix,12pt,prd,superscriptaddress,nofootinbib]{revtex4}
\usepackage{graphicx}
\usepackage{epsfig}
\usepackage{bm}
\usepackage{amsfonts}
\usepackage{subfigure}

\def\bea{\begin{eqnarray}}
\def\eea{\end{eqnarray}}

\begin{document}
\begin{center}
\LARGE { \bf  Warm Gauge-Flation
  }
\end{center}
\begin{center}
{\bf M. R. Setare\footnote{rezakord@ipm.ir} \\  V. Kamali\footnote{vkamali1362@gmail.com}}\\
 { Department of Science, University of Kurdistan,\\
Sanandaj, IRAN.}
 \\
 \end{center}
\vskip 3cm

\begin{center}
{\bf{Abstract}}\\
Non-abelian gauge field inflation is studied in the context of warm inflation scenario. We introduce this scenario as a mechanism that gives an end for
gauge-flation model. Slow-roll parameters and perturbation parameters are presented for this model. We find the general conditions which are required for this model to be realizable in slow-roll approximation. We also develop our model in the context of intermediate and logamediate scenarios which are exact solutions of inflationary field equation in the Einstein theory.  General expressions of slow-roll parameters, tensor-scalar ratio and scalar spectral index  are presented in terms of inflaton field for these two cases. Our model is compatible with
recent observational data from Planck satellite.
\end{center}
\newpage

\section{Introduction}
Inflation model presents  better description of the early phase of the universe. Problems of the Big Bang model, namely the numerical density of monopoles,
flatness, homogeneity and the horizon problem, may find explanations in the framework of the inflationary universe models \cite{1-i}.
Inflation also predicts a mechanism to generate the anisotropy for cosmological microwave background (CMB) and the inhomogeneity for structure formation \cite{2-i}. Cosmic microwave background (CMB) and large scale structure (LSS) observational data denote the isotropic space-time at the background level \cite{2-i,2-m}. Therefore, many models of inflation were proposed versus either single or multi scalar field theories which are isotropic and compatible with the observations of CMB and LSS. These scalar field models have (non)standard kinetic and potential terms which are coupled to gravity. In slow-roll limit, when the kinetic energy of scalar fields is  relatively small compared to the potential energy, the inflation period has been deriven. After this period, in reheating period, the scalar field oscillates around the minimum of the potential while losing its energy to masslese  particles \cite{3-m}. Recently, a new model of inflation was proposed \cite{1} and studied \cite{2} which is generally driven by gauge field theory. This model is widely used in particle physics models but is against the isotropic symmetry of the universe at the background level. This problem could be solved by using three gauge fields which can rotate among each other by non-Abelian gauge transformation SU(2). The rotational symmetry is the global part of SU(2) and isotropy is retained in gauge-flation model \cite{1}.\\
In the warm inflationary models,
radiation production process occurs during inflationary period while
reheating is avoided \cite{3}. Thermal fluctuations may be
obtained during warm inflation. These fluctuations could play a
dominant role to produce initial fluctuations which are necessary
for Large-Scale Structure (LSS) formation. Thus, the density
fluctuation arises from thermal fluctuation rather than quantum fluctuation
\cite{3-i}. Warm inflationary period ends when the universe stops
inflating. After this period, the universe enters radiation
phase smoothly \cite{3}. Finally, remaining inflatons or dominant
radiation fields create the matter components of the universe. Some extensions of this model can be found in Ref.\cite{new}.
The main problem of the
inflation theory is how to attach the universe to the end of the
inflation period. One of the solutions of this problem is
to study the inflation in the context of warm inflation \cite{3}. In
this model, radiation is produced during inflation period where its
energy density is kept nearly constant. This is phenomenologically
fulfilled by introducing the dissipation term $\Gamma$. The study
of warm inflation model as a mechanism that gives an end for
gauge-flation  motivates us to consider the warm gauge-flation model.\\
In the warm inflation, there has to be continuously a particle production.
For this to be possible, the microscopic processes that produce
these particles must occur at a timescale much faster than Hubble
expansion.  Thus, the decay rates $\Gamma_i$ (not to be confused with the
dissipative coefficient) must be larger than $H$. The produced particles need also to be thermalized. Thus, the scattering processes amongst these
produced particles must occur at a rate larger than $H$.  These
adiabatic conditions were outlined since the publication of the early warm inflation
papers, such as Ref. \cite{1-ne}.  More recently,
there has been a considerable explicit calculation from Quantum Field Theory (QFT) that
explicitly computes all the relevant decay and scattering rates
in the warm inflation models \cite{4nn,arj}.\\
In one of the sections of the present work, we would like to consider the warm gauge-flation model in the context of "intermediate inflation". This scenario is one of the exact solutions of inflationary field equation in the Einstein theory with scale factor $a(t)=a_0\exp(At^f)$ ($A>0, 0<f<1$). The study of this model is motivated by string/M theory \cite{4-m}.  By adding the higher order curvature correction, which is proportional to Gauss-Bonnent (GB) term, to Einstein-Hilbert action, one can obtain a free-ghost action \cite{5-m}. Gauss-Bonnent interaction is the leading order in the expansion of inverse string tension, "$\alpha$"  to low-energy string effective action \cite{5-m}. This theory may be applied for black hole solutions \cite{6-m}, acceleration of the late time universe \cite{7-m} and initial singularity problems \cite{8-m}. The GB interaction in $4d$ with dynamical dilatonic scalar coupling leads to an intermediate form of the scale factor \cite{4-m}. Expansion of the universe in the intermediate inflation scenario is slower than standard de sitter inflation with scale factor $a=a_0\exp(H_0 t)$ ($a_0, H_0>0$), but faster than power-low inflation (with the scale factor $a=t^p$ ($p>1$)). Harrison-Zeldovich \cite{10-m}  spectrum of the density perturbation i.e. $n_s=1$ for the intermediate inflation models  driven by scalar field is presented for exact values of parameter $f$ \cite{11-m}.\\
On the other hand, we will also study our model in the context of "logamediate inflation" with the scale factor $a(t)=a_0 \exp(a[\ln t]^{\lambda})$ ($\lambda >1, A>0$) \cite{12-m}. This model is converted to power-law inflation for $\lambda=1$ cases. This scenario is applied in a number of scalar-tensor theories \cite{13-m}. The study of the logamediate scenario is motivated by imposing weak general conditions on the cosmological models which have indefinite expansions \cite{12-m}. The effective potential of the logamediate model has been considered in dark energy models \cite{14-m}. These forms of potentials are also used in supergravity, Kaluza-Klein theories and super-string models \cite{13-m,15-m}. For logamediate models, the power spectrum could be either red or blue tilted \cite{16-m}. In Ref.\cite{12-m}, we can find eight possible asymptotic scale factor solutions for cosmological dynamics. Three of these solutions are non-inflationary scale factor, another three  solutions gives power-low, de sitter and intermediate scale factors. Finally, two cases of these solutions have asymptotic expansions with logamediate scale factor.
\section{Warm inflation driven by non-abelian gauge fields}
Gauge-flation model in a flat-space Friedmann-Robertson-Walker (FRW) background is described by an effective Lagrangian \cite{1}
\begin{eqnarray}\label{1}
\mathcal{L}=\sqrt{-g}(-\frac{R}{2}-\frac{1}{4}F^{a}_{\mu\nu}F^{\mu\nu}_a+\frac{\kappa^2}{384}(\epsilon^{\mu\nu\lambda\sigma}F^{a}_{\mu\nu}F^{a}_{\lambda\sigma})^2)
\end{eqnarray}

where $8\pi G=M_p^{-2}=1$, $\epsilon^{\mu\nu\rho\sigma}$ is an antisymmetric tensor and
\begin{eqnarray}\label{2}
F^{a}_{\mu\nu}=\partial_{\mu}A^{a}_{\nu}-\partial_{\nu}A^{a}_{\mu}-g\epsilon^{a}_{bc}A^{b}_{\nu}A^{c}_{\mu}
\end{eqnarray}
where $\epsilon^{a}_{bc}$ is also antisymmetric tensor. It was shown that $F^4$ term in Eq.(\ref{1}) can be derived by integrating out an axion field in Chromo-Natural inflation \cite{2}. To obtain isotropy symmetry of space-time, we can introduce the effective inflaton field as \cite{1}:

\begin{eqnarray}\label{3}
A^{a}_{\mu}=\{^{\phi(t)\delta^a_i~~~\mu=i}_{0~~~~~~~~\mu=0}=\{^{a(t)\psi(t)\delta^a_i~~~\mu=i}_{0~~~~~~~~~~~~\mu=0}
\end{eqnarray}
Using the above ansatz, we could find a reduced effective Lagrangian from Eq.(\ref{1})
\begin{eqnarray}\label{4}
\mathcal{L}_{red}=\frac{3}{2}(\frac{\dot{\phi}^2}{a^2}-\frac{g^2\phi^4}{a^4}+\kappa\frac{g^2\dot{\phi}^2\phi^4}{a^6})
\end{eqnarray}
Pressure and energy density have the following forms
\begin{eqnarray}\label{5}
P_{\phi}=\frac{1}{3}\rho_{YM}-\rho_{F^4}~~~~~~~~\rho_{\phi}=\rho_{YM}+\rho_{F^4}
\end{eqnarray}
where
\begin{eqnarray}\label{6}
\rho_{YM}=\frac{3}{2}(\frac{\dot{\phi}^2}{a^2}+\frac{g^2\phi^4}{a^4})~~~~~~~~~~\rho_{F^4}=\frac{3}{2}\kappa\frac{g^2\dot{\phi}^2\phi^4}{a^6}
\end{eqnarray}
$F^4$ term has been chosen because the contribution of this term to presser and energy density leads to $P=-\rho$ which gives the inflationary dynamic. The gauge field without $F^4$ term has the equation of state of radiation ($P=\frac{\rho}{3}$) and could not be source of inflation.
The dynamics of phenomenological warm gauge-flation in spatially flat FRW model  is described by these equations
\begin{eqnarray}\label{7}
H^2=\frac{1}{3}(\rho_{\phi}+\rho_{\gamma})\\
\nonumber
\dot{\rho}_{\phi}+3H(\rho_{\phi}+P_{\phi})=-\Gamma\frac{\dot{\phi}^2}{a^2}\\
\nonumber
\dot{\rho}_{\gamma}+4H\rho_{\gamma}=\Gamma\frac{\dot{\phi}^2}{a^2}
\end{eqnarray}
where $\rho_{\gamma}$ is energy density of the radiation, $H$ is Hubble parameter and
$\Gamma$ is the dissipative coefficient.
From  Eqs.(\ref{5}) and (\ref{6}), the  equation of motion is reduced to
\begin{eqnarray}\label{8}
(1+\kappa g^2\frac{\phi^4}{a^4})\frac{\ddot{\phi}}{a}+(1+r-3\kappa g^2\frac{\phi^4}{a^4})H\frac{\dot{\phi}}{a}+(1+\frac{\kappa \dot{\phi}^2}{a^2})\frac{2g^2\phi^3}{a^3}=0
\end{eqnarray}
where $r=\frac{\Gamma}{3H}$.
In the above equations dots "." mean derivative with
respect to cosmic time. During inflation epoch, the energy density
$\rho_{\phi}$ is of the order of potential energy density $\rho_{F^4}$ ($\rho_{\phi}\sim V$) and
dominates over the radiation energy $\rho_{\phi}>\rho_{\gamma}$.
Using slow-roll approximation when
$\ddot{\phi}\ll(3H+\Gamma)\dot{\phi}$ \cite{3} and when inflation
radiation production is quasi-stable ($\dot{\rho}_{\gamma}\ll
4H\rho_{\gamma}$, $\dot{\rho}_{\gamma}\ll\Gamma\frac{\dot{\phi}^2}{a^2}$), we can see the
dynamic equations (\ref{7}) and (\ref{8}) are reduced to
\begin{eqnarray}\label{9}
H^2=\frac{1}{72\kappa g^2}\frac{\Gamma^2}{\psi^2}~~~~~~~~~~~~~~~~~~~~~~~~~~~~~~~~~~~~~~~~~~~~\\
\nonumber
\dot{H}+\frac{r}{2}H^2\psi^2+g^2\psi^4=0,~~~~~~~\psi^2\simeq-\frac{6}{\Gamma}\frac{\dot{H}}{H}~~~~~~~~~~\\
\nonumber
\frac{\dot{\phi}}{a}=-\frac{\Gamma}{6\kappa g^2\psi^2}~~~~~~~~~~~~~~~~~~~~~~~~~~~~~~~~~~~~~~~~~~~~~\\
\nonumber
\rho_{\gamma}=\frac{\Gamma}{4H}\frac{\dot{\phi}^2}{a^2}\simeq\frac{\Gamma }{4H}H^2\psi^2=\frac{\Gamma^2}{24(2\kappa g^2)^{\frac{1}{2}}}=\sigma T_r^4~~~~~~~
\end{eqnarray}
where $\sigma$ is Stefan-Boltzmann constant and $T_r,$ is the temperature of thermal bath.
Slow-roll parameters of the warm gauge-flation are
\begin{eqnarray}\label{10}
\epsilon=-\frac{1}{H}\frac{d}{dt}\ln H=-\frac{\Gamma'}{\Gamma}-2\frac{\psi'}{\psi}=\frac{1}{2}r\psi^2\\
\nonumber
\eta=\frac{1}{2}\epsilon-\frac{\dot{\epsilon}}{H\epsilon}=\frac{1}{2}r\psi^2-\frac{r'}{r}-2\frac{\psi'}{\psi}~~~~~~~\\
\nonumber
\delta=-\frac{\dot{\psi}}{H\psi}=-\frac{\psi'}{\psi}~~~~~~~~~~~~~~~~~~~~~~~~~~~~
\end{eqnarray}
where primes ($'$) denote a derivative with respect to the number of e-folds $N$ ($dN=Hdt$).
The scalar field at the beginning of inflation is given by $\psi_i^2=\frac{2}{r}$ when $\epsilon=1$.
From Eqs.(\ref{9}) and (\ref{10}), we could find a relation between
$\rho_{\phi}$ and $\rho_{\gamma}$
\begin{eqnarray}\label{11}
\rho_{\gamma}=\frac{\epsilon}{2}\rho_{\phi}
\end{eqnarray}
Condition of inflation epoch ($\ddot{a}<1$) may be obtained by inequality  $\epsilon<1$  \cite{1}.
\begin{eqnarray}\label{12}
\rho_{\phi}>2\rho_{\gamma}
\end{eqnarray}
 Warm inflation epoch ends when $\rho_{\phi}=2\rho_{\gamma}$. The number of e-folds
has the following form
\begin{eqnarray}\label{13}
N=\int Hdt=\frac{1}{\sqrt{72\kappa g^2}}\int_{\phi_{*}}^{\phi_f}\frac{\Gamma a}{\phi} dt
\end{eqnarray}
where $f$ denotes the end of inflation and $*$ denotes the epoch when the cosmological
scale exits the horizon. \\
Now we study the perturbation of our model at the smallest level in spatially flat FRW background. We will consider the perturbation theory in isotropic universe using variation of inflaton $\psi$. In warm inflation scenario, the variation of field is presented by thermal fluctuation. In non-warm scenarios quantum fluctuation predicts \cite{1,2}
\begin{eqnarray}\label{14}
\langle\delta\psi\rangle_{quantum}=\frac{H^2}{2\pi}
\end{eqnarray}
but in warm inflation model the thermal fluctuation provides \cite{3,3-i}
\begin{eqnarray}\label{15}
\langle\delta\psi\rangle_{thermal}=(\frac{\Gamma H T^2}{(4\pi)^3})^{\frac{1}{4}}
\end{eqnarray}
Scalar and tensor perturbation emerging during inflation epoch will be studied in warm gauge-flation model. These perturbations may leave an imprint in the CMB anisotropy  and on the LSS \cite{4,5}. Power spectrum and a spectral index, are characteristics of each fluctuation: $\Delta_T^2(k)$ and $n_T$ for the tensor perturbation, $\Delta_R^2(k)$ and $n_R$ for the scalar perturbation. In warm and cool inflation models, scalar power spectrum is given by \cite{3}
\begin{eqnarray}\label{16}
\Delta_R^2(k)=(\frac{H}{\dot{\phi}}\langle\delta\phi\rangle)^2
\end{eqnarray}
where $k$ is co-moving wavenumber.
In our model, the power-spectrum of the scalar perturbation is presented from Eqs.(\ref{9}) and (\ref{15})
\begin{eqnarray}\label{17}
\Delta_{R}^2(k)=\frac{1}{(4\pi)^{\frac{3}{2}}(72\kappa g^2)^{\frac{1}{4}}}\frac{\Gamma T}{\psi^{\frac{5}{2}}}
\end{eqnarray}
The largest value of the density perturbation is produced when $\psi=\psi_i$ \cite{6}. The scalar spectral index of our model is presented by
\begin{eqnarray}\label{18}
n_s-1=-\frac{d\ln \Delta_{R}^2(k)}{d\ln k}=\frac{3}{2}\epsilon-\eta
\end{eqnarray}
The Planck measurement constraints this parameter as \cite{2-i}:
\begin{eqnarray}\label{}
n_s=0.96\pm0.0073
\end{eqnarray}
In the warm inflation scenario, the thermal fluctuations are considered instead of the quantum fluctuations that generate  the scalar perturbations. Therefore,  the density fluctuation of the scalar perturbation is modified while the tensor perturbation shows the same spectrum as in the usual non-warm inflation \cite{16}
\begin{eqnarray}\label{19}
\Delta^2_T=\frac{2 H^2}{\pi^2}=\frac{1}{36\kappa g^2\pi^2}\frac{\Gamma^2}{\psi^2}
\end{eqnarray}
The spectral index $n_T$ may be found as
\begin{eqnarray}\label{}
n_T=-2\epsilon
\end{eqnarray}
From observational data,  $\Delta^2_T$ could not be constrained directly, but the tensor-scalar ratio could  be constrained
\begin{eqnarray}\label{21}
R=\frac{\Delta^2_T}{\Delta^2_R}=\frac{2^{\frac{1}{4}}(4\pi)^{\frac{3}{2}}}{(36\kappa g^2)^{\frac{3}{4}}}\frac{\Gamma\psi^{\frac{1}{2}}}{T}
\end{eqnarray}
The Planck measurement constraints this parameter as \cite{2-i,2-m}:
\begin{eqnarray}\label{22}
 R \leq 0.11
\end{eqnarray}
\section{Intermediate inflation}
Intermediate inflation will be studied in this section, where the scale factor of this model is given by
\begin{eqnarray}\label{23}
a=a_0\exp(At^f)~~~~~~0<f<1
\end{eqnarray}
where $A$ is a positive constant. Using the above equation, we obtain the number of e-fold  as
\begin{eqnarray}\label{24}
N=\int_{t_1}^{t} H dt=A(t^{f}-t_1^{f})
\end{eqnarray}
where $t_1$ is the beginning time of the inflation. From Eqs.(\ref{9}) and (\ref{23}), we could find the Hubble parameter and scalar field as
\begin{eqnarray}\label{25}
H(\psi)=fA(\frac{\psi^{-2}}{\beta})^{f-1}\\
\nonumber
\psi^{-2}=\beta t~~~~~~~~~
\end{eqnarray}
where $\beta=\frac{\Gamma}{6(1-f)}$ and $\Gamma=const$. Important slow-roll parameters $\epsilon$ and $\eta$ are given by
\begin{eqnarray}\label{26}
\epsilon=\frac{1-f}{fA}(\frac{\psi^{-2}}{\beta})^{-f}\\
\nonumber
\eta=\frac{2-f}{fA}(\frac{\psi^{-2}}{\beta})^{-f}
\end{eqnarray}
respectively. The number of e-fold between two fields $\psi_1$ and $\psi$ is presented, using Eq.(\ref{24}), by
\begin{eqnarray}\label{27}
N=A((\frac{\psi^{-2}}{\beta})^{f}-(\frac{\psi^{-2}_1}{\beta})^{f})=A((\frac{\psi^{-2}}{\beta})^{f}-\frac{1-f}{fA})
\end{eqnarray}
\begin{figure}[h]
\centering
  \includegraphics[width=10cm]{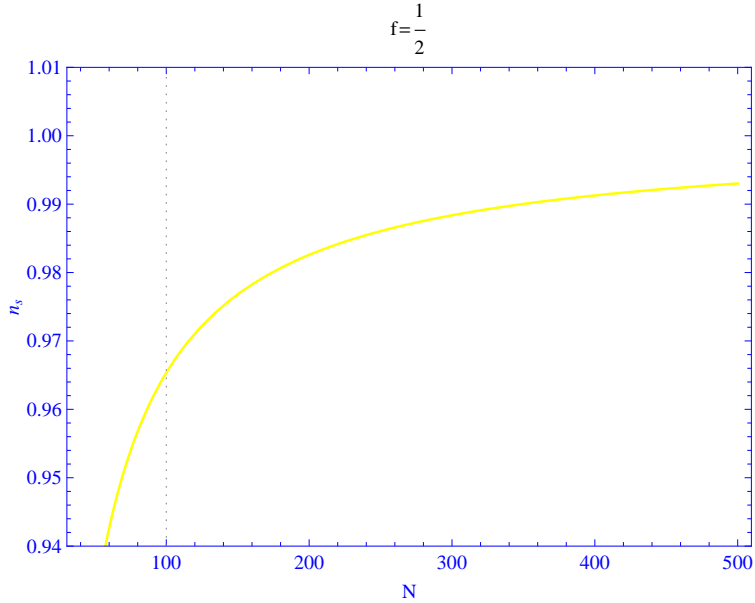}
  \caption{In this graph, we plot the spectral index $n_s$   in terms of the number of e-folds $N$ for intermediate scenario ($A=1$, $f=\frac{1}{2}, \Gamma\propto\frac{1}{T^{\frac{2}{3}}}$).}
 \label{fig:F3}
\end{figure}
At the beginning of the inflation epoch where $\epsilon=1,$ we find the scalar field in terms of constant parameters of the model
\begin{eqnarray}\label{28}
\psi^{-2}_1=\beta(\frac{1-f}{fA})^{\frac{1}{f}}
\end{eqnarray}
By using the above equations, we may find the scalar field $\psi(t)$ in terms of the number of e-folds
\begin{eqnarray}\label{29}
\psi^{-2}=\beta(\frac{N}{A}+\frac{1-f}{fA})^{\frac{1}{f}}
\end{eqnarray}
\begin{figure}[h]
\centering
  \includegraphics[width=10cm]{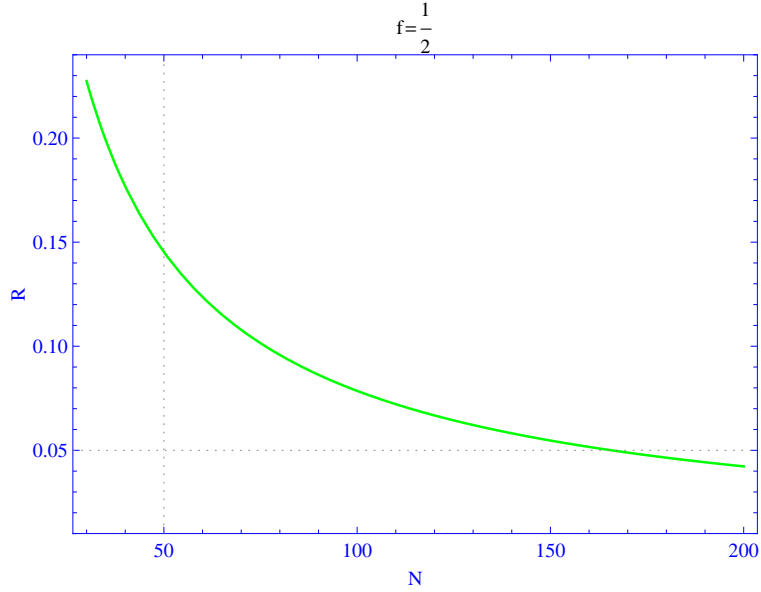}
  \caption{In this graph, we plot the scalar-tensor ratio $R$   in term of the number of e-folds $N$ for intermediate scenario ($A=1,f=\frac{1}{2}, \Gamma\propto\frac{1}{T^{\frac{2}{3}}}$).}
 \label{fig:F3}
\end{figure}
Perturbation parameters versus the scalar field $\psi,$ and constant parameters of intermediate scenario are presented for warm gauge-flation model.
In the slow-roll limit, the power-spectrum of scalar perturbation could be found, using Eqs.(\ref{15}), (\ref{16}) and (\ref{29}), as
\begin{eqnarray}\label{30}
\Delta_R^2=T(\frac{\Gamma fA}{(4\pi)^3})^{\frac{1}{2}}\psi^{-2}(\frac{\psi^{-2}}{\beta})^{\frac{f-1}{2}}\\
\nonumber
=T(\frac{\Gamma fA}{(4\pi)^3})^{\frac{1}{2}}\beta(\frac{N}{A}+\frac{1-f}{fA})^{\frac{f+1}{2f}}
\end{eqnarray}
Another important perturbation parameter is spectral index $n_s$ which is given by
\begin{eqnarray}\label{31}
n_s-1=-\frac{d\ln\Delta^2_R}{dN}=-\frac{f+1}{2fA}(\frac{N}{A}+\frac{1-f}{fA})^{-1}
\end{eqnarray}
In Fig.(1),  the spectral index $n_s$  versus the number of e-folds is plotted (where $f=\frac{1}{2}$).
 Our model is compatible with observational data \cite{2-i}, ($N\simeq 70$ case  leads to $n_s\simeq 0.96$).
Tensor power spectrum and its spectral index are given by
\begin{eqnarray}\label{32}
\Delta^2_T=\frac{2f^2A^2}{\pi^2}(\frac{\psi^{-2}}{\beta})^{2(f-1)}=(\frac{N}{A}+\frac{1-f}{fA})^{\frac{2f-2}{f}}\\
\nonumber
n_T=\frac{2f-2}{fA}(\frac{\psi^{-2}}{\beta})^{-f}~~~~~~~~~~~~~~~~~~~~~~~~~
\end{eqnarray}
Tensor-scalar ratio has the following form
\begin{eqnarray}\label{33}
R=\frac{2}{T\Gamma^{\frac{1}{2}}}\frac{(fA)^{\frac{3}{2}}}{\pi^{\frac{1}{2}}}\psi^2(\frac{\psi^{-2}}{\beta})^{\frac{3f-3}{2}}~~~~~~~~~~~~~~~~~~\\
\nonumber
=\frac{2}{T\Gamma^{\frac{1}{2}}}\frac{(fA)^{\frac{3}{2}}}{\pi^{\frac{1}{2}}}\beta(\frac{N}{A}+\frac{1-f}{fA})^{\frac{3f-5}{2f}}~~~~\\
\end{eqnarray}
Tensor-scalar ratio $R$ in terms of the number of e-folds is plotted in Fig.(2). Standard case $N\geq 70$, leads to $0.R<0.11,$
 which agrees  with observational data \cite{2-i,2-m}.
\section{Logamediate inflation}
In this section, we study warm gauge field logamediate inflation where the scale factor has the following form
\begin{eqnarray}\label{34}
a(t)=a_0\exp(A[\ln t]^{\lambda}) ~~~~~~\lambda>1,
\end{eqnarray}
$A$ is a constant parameter. By using the above equation, we may find the number of e-folds as
\begin{eqnarray}\label{35}
N=\int_{t1}^{t} H dt=A[(\ln t)^{\lambda}-(\ln t_1)^{\lambda}]
\end{eqnarray}
where $t_1$ denotes the beginning time of inflation epoch. Using Eqs.(\ref{9}) and (\ref{34}), we may find the inflaton $\psi$
\begin{eqnarray}\label{36}
\psi=\Xi(t)
\end{eqnarray}
where $\Xi(t)=[\frac{r\lambda A(\ln t)^{\lambda-1}}{t}(-1+\sqrt{1+\frac{16g^2 t^2}{r^2(\ln t)^{3\lambda-3}}})]^{\frac{1}{2}}$. The Hubble parameter in terms of scalar field $\psi$ is presented as:
\begin{eqnarray}\label{37}
H=\lambda A\frac{(\ln \Xi^{-1}(\psi))^{\lambda-1}}{\Xi^{-1}(\psi)}
\end{eqnarray}
\begin{figure}[h]
\begin{minipage}[b]{1\textwidth}
\subfigure[\label{fig1a} ]{ \includegraphics[width=.37\textwidth]%
{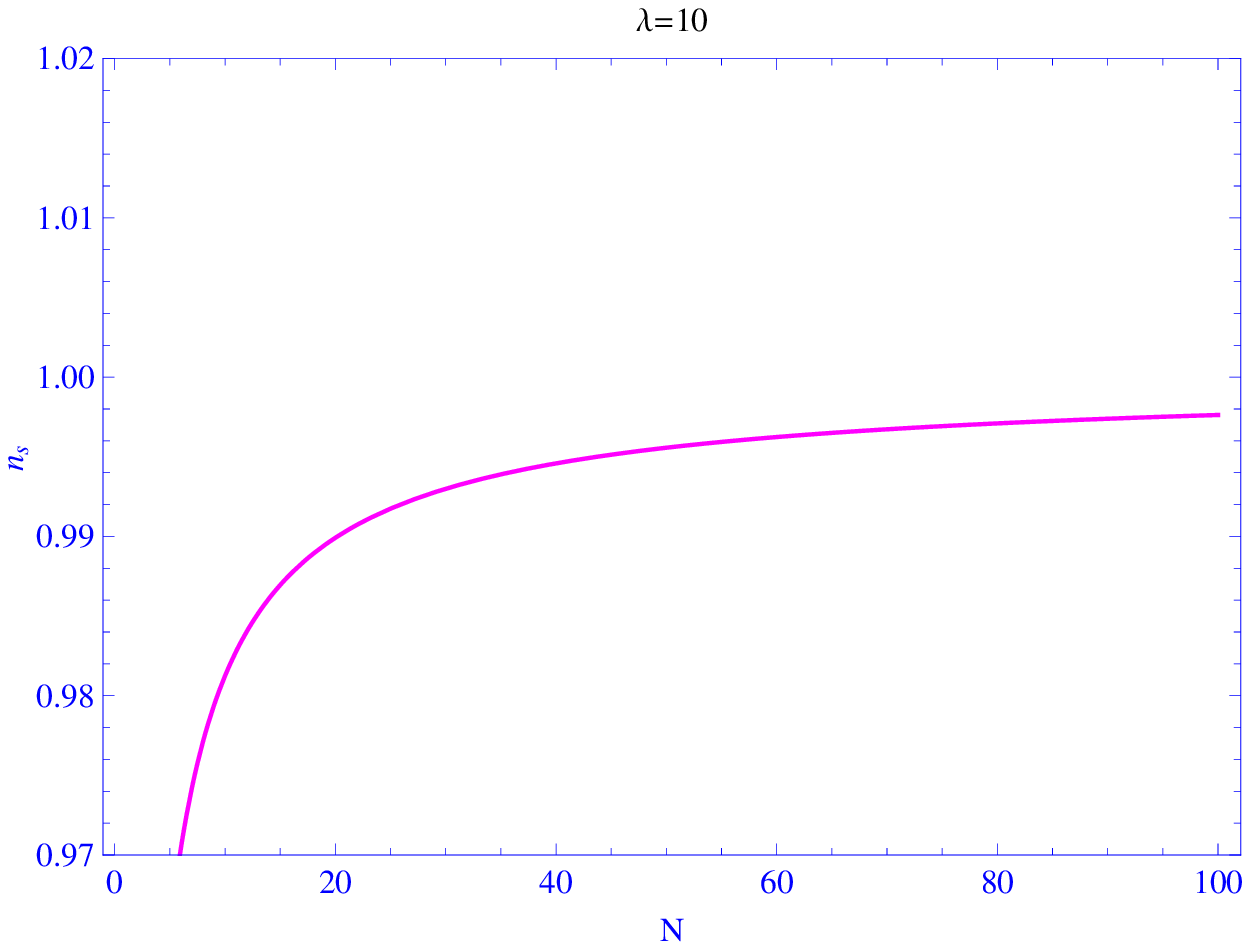}} \hspace{.2cm}
\subfigure[\label{fig1b} ]{ \includegraphics[width=.37\textwidth]%
{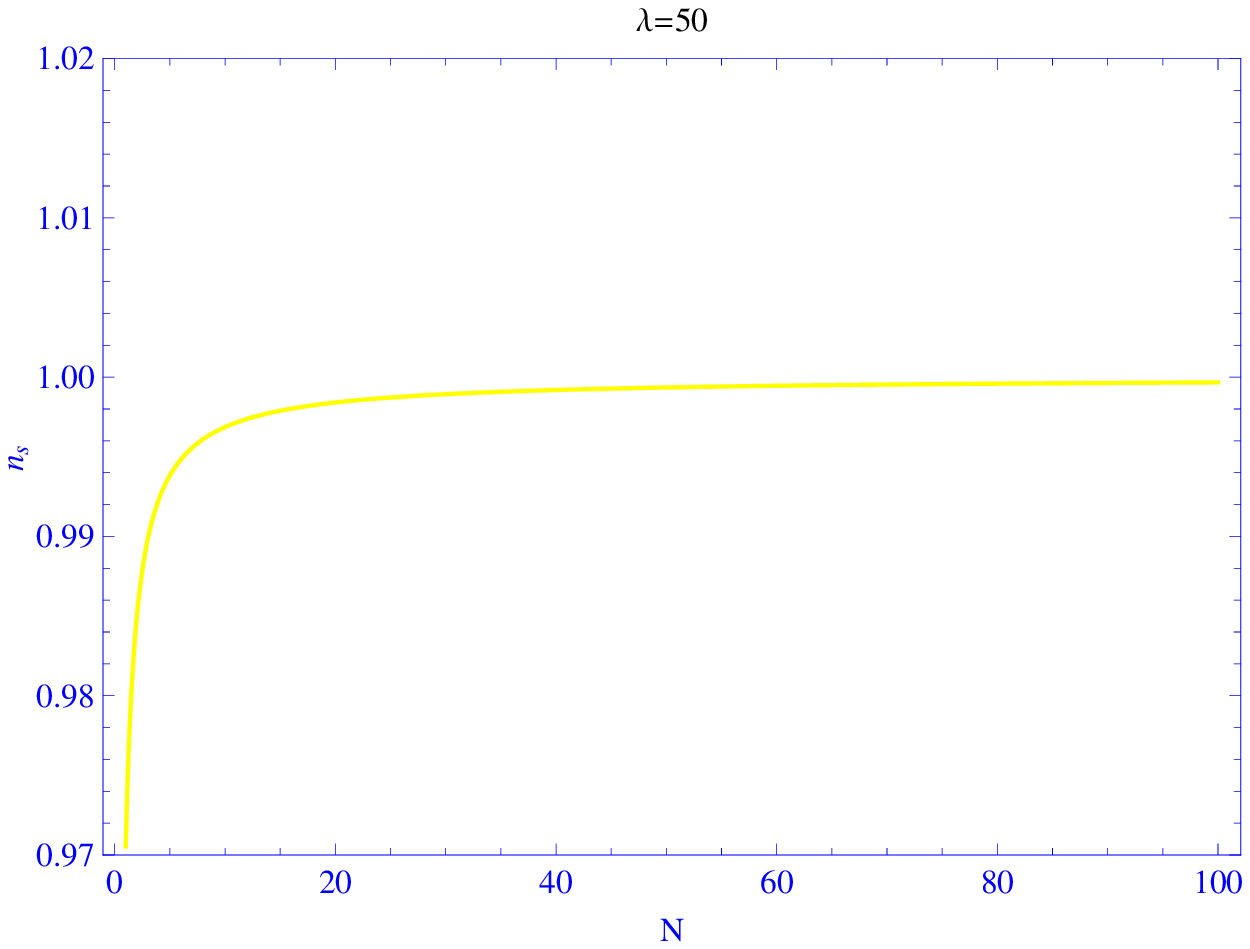}}
\end{minipage}
\caption{ Spectral index $n_s$ in terms of the number of e-folds $N$: (a) for $\lambda=10$ and (b) for $\lambda=50$ (where   $\Gamma\propto\frac{1}{T^{\frac{2}{3}}}$ ).}
\end{figure}
The slow-roll parameters of the model in this case are given by
\begin{eqnarray}\label{38}
\epsilon=\frac{[\ln \Xi^{-1}(\psi)]^{1-\lambda}}{\lambda A}\\
\nonumber
\eta=\frac{2[\ln \Xi^{-1}(\psi)]^{1-\lambda}}{\lambda A}
\end{eqnarray}
The number of e-folds between two fields $\psi_1$ and $\psi(t)$ may be determined (From Eqs.(\ref{35}) and (\ref{36})).
\begin{eqnarray}\label{39}
N=A((\ln \Xi^{-1}(\psi))^{\lambda}-(\ln \Xi^{-1}(\psi_1))^{\lambda})=A((\ln \Xi^{-1}(\psi))^{\lambda}-(\lambda A)^{\frac{\lambda}{1-\lambda}})
\end{eqnarray}
where $\psi_1$ is the inflaton at the beginning of the inflation epoch where $\epsilon=1$. Inflaton field in the inflation period may be obtained in terms of the number of e-folds (From the above equation).
\begin{eqnarray}\label{40}
\psi=\Xi[\exp([\frac{N}{A}+(\lambda A)^{\frac{\lambda}{1-\lambda}}]^{\frac{1}{\lambda}})]
\end{eqnarray}
Power spectrum and tensor-scalar ratio in this case are given by
\begin{eqnarray}\label{41}
\Delta^2_R=T(\frac{\Gamma \lambda A}{(4\pi)^3})^{\frac{1}{2}}\frac{(\ln\Xi^{-1}(\psi))^{\frac{\lambda-1}{2}}}{\psi^2\Xi^{-1}(\psi)}~~~~~~~~~~~~~~~~~~~~~~~~~~~~\\
\nonumber
=T(\frac{\Gamma \lambda A}{(4\pi)^3})^{\frac{1}{2}}[\frac{N}{A}+(\lambda A)^{\frac{\lambda}{1-\lambda}}]^{\frac{\lambda-1}{2\lambda}}~~~~~~~~~~~~~~~~~~~~~~~~~~~~\\
\nonumber
\times\exp(-[\frac{N}{A}+(\lambda A)^{\frac{\lambda}{1-\lambda}}]^{\frac{1}{\lambda}})\Xi^2[\exp([\frac{N}{A}+(\lambda A)^{\frac{\lambda}{1-\lambda}}]^{\frac{1}{\lambda}})]\\
\nonumber
R=\frac{2(\lambda A)^{\frac{3}{2}}}{\pi^2 T}(\frac{(4\pi)^3}{\Gamma})^{\frac{1}{2}}\frac{(\ln\Xi^{-1})^{\frac{3\lambda-3}{2}}}{(\Xi^{-1})^{\frac{3}{2}}}\psi^2~~~~~~~~~~~~~~~~~~~~~~
\end{eqnarray}
Spectral indices for our model have the following forms
\begin{eqnarray}\label{42}
n_s-1=\frac{3}{2}\epsilon-\eta=-\frac{1}{2}\frac{[\ln\Xi^{-1}(\psi)]^{1-\lambda}}{\lambda A}\\
\nonumber
=-\frac{1}{2\lambda A}[\frac{N}{A}+(\lambda A)^{\frac{\lambda}{1-\lambda}}]^{\frac{1-\lambda}{\lambda}}~~~~~\\
\nonumber
n_T=-2\frac{(\ln\Xi^{-1})^{1-\lambda}}{\lambda A}~~~~~~~~~~~~~~~~~~~~~
\end{eqnarray}
In Fig.(3), the spectral index $n_s$ in terms of the number of e-folds is plotted  (for $\lambda=10$, $\lambda=50$,  cases). We can see the small values of number of e-folds are assured for large values of $\lambda$ parameter.
We could find the tensor-scalar ratio in terms of the number of e-folds and spectral index $n_s$
\begin{eqnarray}\label{43}
R=\frac{2(\lambda A)^{\frac{3}{2}}}{\pi^2 T}(\frac{(4\pi)^3}{\Gamma})^{\frac{1}{2}}\Xi^2[\exp(\frac{N}{A}+(\lambda A)^{\frac{\lambda}{1-\lambda}})^{\frac{1}{\lambda}}]~~~~~~~~~~~~~~\\
\nonumber
\times(\frac{N}{A}+(\lambda A)^{\frac{\lambda}{1-\lambda}})^{\frac{3\lambda-3}{2\lambda}}\exp[-\frac{3}{2}(\frac{N}{A}+(\lambda A)^{\frac{\lambda}{1-\lambda}})^{\frac{1}{\lambda}}]~~~~~~~\\
\end{eqnarray}
\begin{figure}[h]
\begin{minipage}[b]{1\textwidth}
\subfigure[\label{fig1a} ]{ \includegraphics[width=.37\textwidth]%
{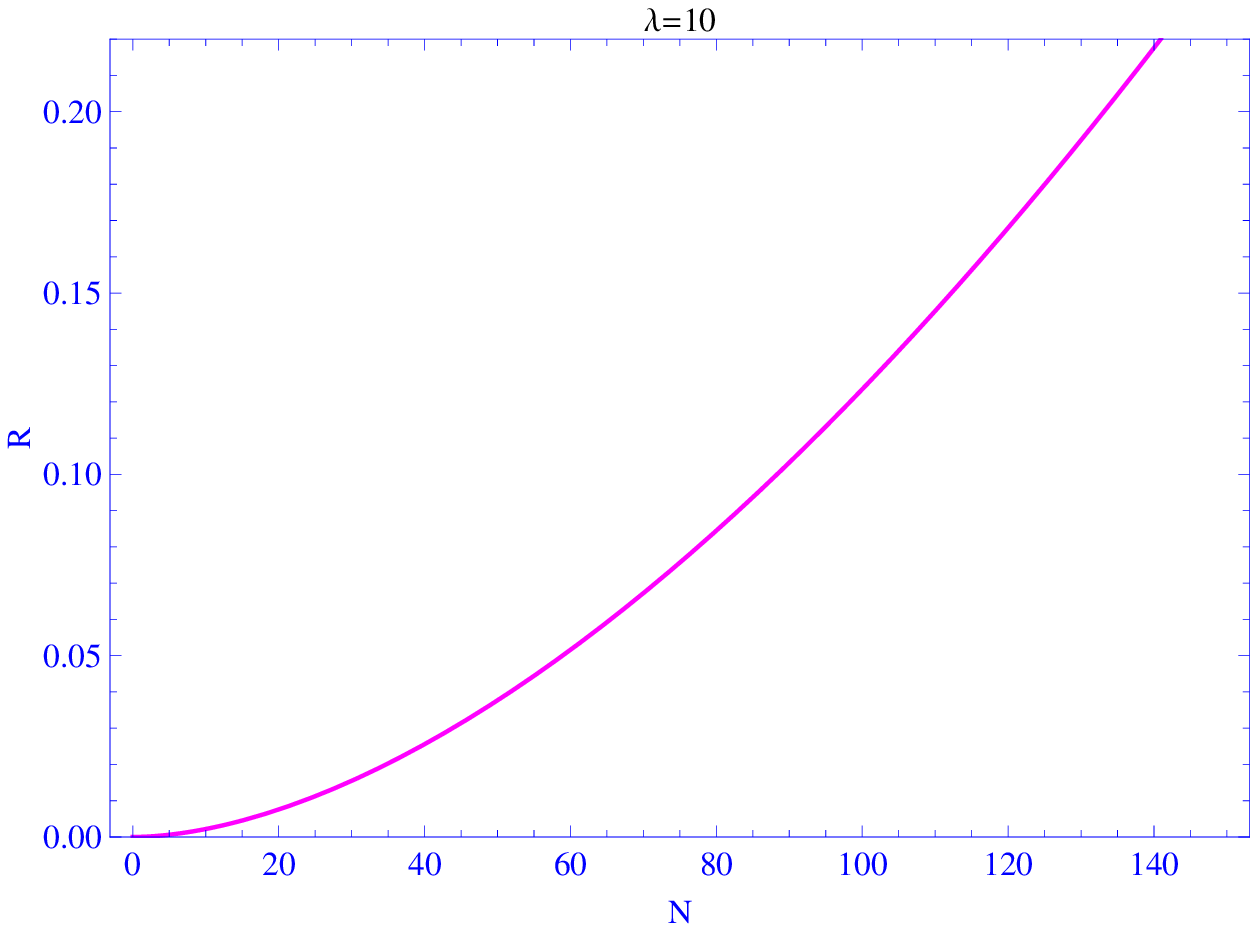}} \hspace{.2cm}
\subfigure[\label{fig1b} ]{ \includegraphics[width=.37\textwidth]%
{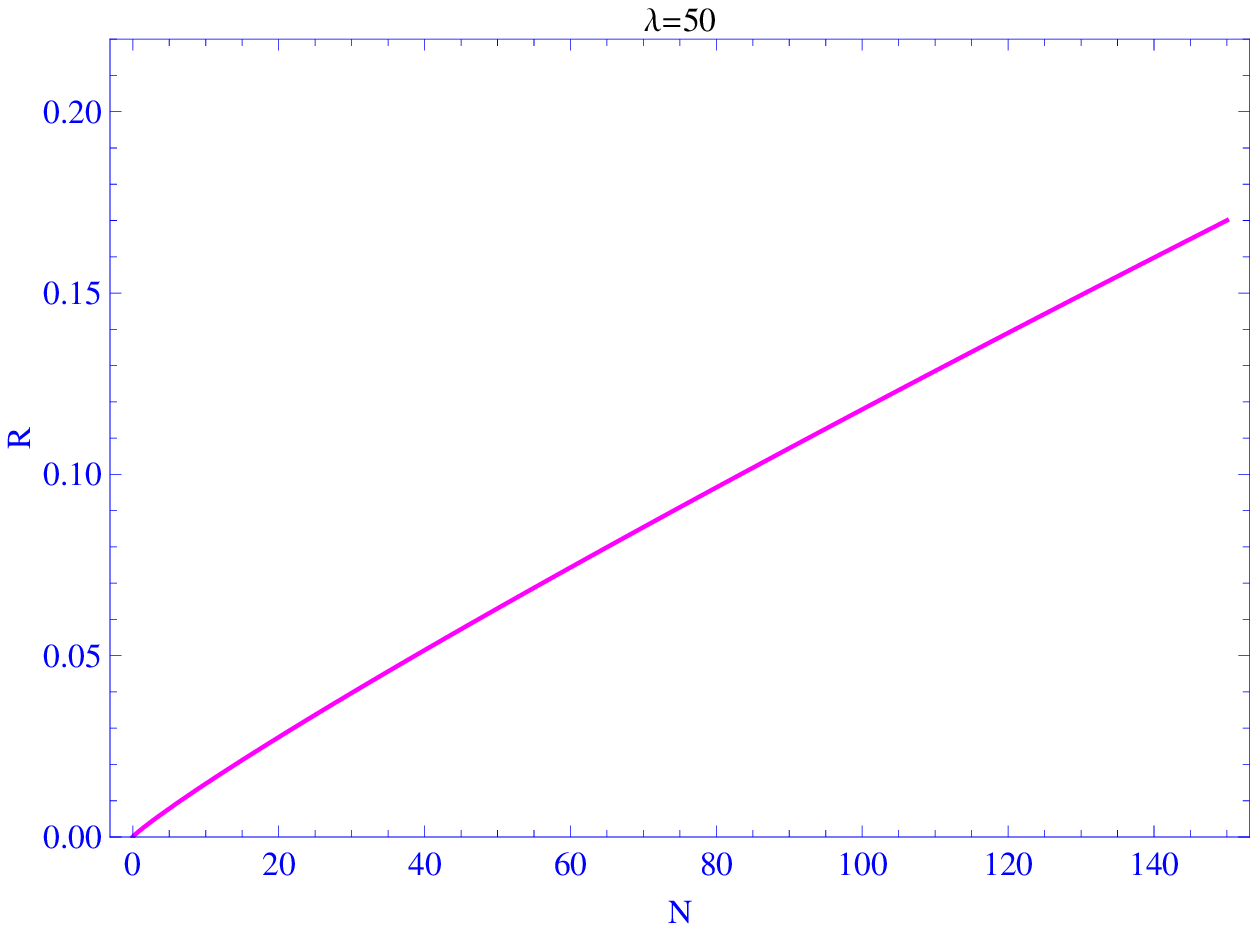}}
\end{minipage}
\caption{ Tensor-scalar ratio  in term of number of e-folds: (a) for $\lambda=10$ and (b) for $\lambda=50$ (where $\Gamma\propto\frac{1}{T^{\frac{2}{3}}}$.}
\end{figure}
In Fig.(4), the tensor-scalar ratio  versus the number of e-folds is plotted  for $\lambda=10$, $\lambda=50$. We find the model to be compatible with observational data \cite{2-i,2-m} ( $60<N<100,$  leads to $R<0.11$).
\section{Conclusion}
In this work we have studied the gauge-field inflation model in the context of warm inflation.
The main problem of inflation theory is how to attach the universe to the end of the
inflation period. The study of phenomenological warm inflation scenario as a mechanism that gives an end for
inflation models  motivates us to consider the  gauge-flation model using warm inflation theory. We have found the general conditions which are required for our model to be realizable in slow-roll limit. The model is compatible with Planck observational data. We have developed our model in the intermediate and logamediate scenarios. In the intermediate scenario, the numerical study for $f=\frac{1}{2}$ case leads to best compatibility  with observational data ($N>70,$ leads to $R<0.11,$ and $N\simeq 70,$ leads to $n_s\simeq 0.96$ ). In the logamediate scenario, we have studied $\lambda=10,$ and $\lambda=50,$ cases. In $\lambda=50$ case, where $60<N<100$ we have found Planck result, $R<0.11$.

\end{document}